\documentclass[a4paper,12pt]{article}
\usepackage{jcappub}
\usepackage{amsfonts}
\usepackage{amsmath}
\usepackage{amssymb}
\usepackage[T1]{fontenc}
\usepackage{graphicx}
\providecommand{\U}[1]{\protect\rule{.1in}{.1in}}
\begin{document}

\title{Dark Matter and Strong Electroweak Phase Transition in a Radiative Neutrino
Mass Model}
\author{Amine Ahriche and Salah Nasri}

\affiliation[a]{Department of Physics, University of Jijel, PB 98
Ouled Aissa, DZ-18000 Jijel, Algeria.} \emailAdd{aahriche@ictp.it}

\affiliation[b]{Physics Department, UAE University, POB 17551, Al
Ain, United Arab Emirates.} \emailAdd{snasri@uaeu.ac.ae}

\abstract{We consider an extension of the standard model (SM) with
charged singlet scalars and right handed (RH) neutrinos all at the
electroweak scale. In this model, the neutrino masses are generated
at three loops, which provide an explanation for their smallness,
and the lightest RH neutrino, $N_{1}$, is a dark matter candidate.
We find that for three generations of RH neutrinos, the model can be
consistent with the neutrino oscillation data, lepton flavor
violating processes, $N_{1}$ can have a relic density in agreement
with the recent Planck data, and the electroweak phase transition
can be strongly first order. We also show that the charged scalars
may enhance the branching ratio $h\rightarrow\gamma\gamma$, where as
$h\rightarrow\gamma Z$ get can get few percent suppression. We also
discuss the phenomenological implications of the RH neutrinos at the
collider.}

\maketitle

\section{Introduction}

There are three concrete evidences for Physics beyond the standard model (SM):
(i) non zero neutrino masses, (ii) the existence of dark matter (DM), and
(iii) the observation of matter anti matter asymmetry of the universe.
However, most of the SM extensions make no attempt to address these three
puzzles within the same framework. For instance, in the minimal supersymmetric
standard model (MSSM), the lightest supersymmetric particle (LSP) is a
candidate for DM and, in principle, has the necessary ingredients to generate
the baryon asymmetry of the universe (BAU), but it does not provide an
explanation for why neutrino masses are tiny. Moreover, direct searches for
supersymmetric particles have yielded null results so far. An interesting
class of models which has a DM candidate and can, in principle, generate the
BAU is the so called inert doublet model \cite{Ma, Barbieri, cline}. Another
popular extension of the SM, is introducing very heavy right-handed (RH)
neutrinos ($m_{N}\geq10^{8}~\mathrm{GeV}$, where small neutrino masses are
generated via the see-saw mechanism \cite{seesaw}, and the BAU is produced via
leptogenesis \cite{FY}. Unfortunately, such heavy particles decouple from the
effective low energy theory and can not be tested at collider experiments. In
addition, for $m_{N}$ heavier than $10^{7}$ \textrm{GeV}, the Dirac neutrino
mass term induces large corrections to the Higgs mass, which can destabilize
the electroweak vacuum \cite{hierarchy}.

Another possible way to understand the smallness of neutrino masses is to
generate them radiatively. The famous example is the so-called Zee model
\cite{Zee}, where one augments the scalar sector of the SM with a Higgs
doublet, and a charged field which transforms as a singlet under $SU(2)_{L}$,
which leads to non zero neutrino mass at one loop level. However, the solar
mixing angle comes out to be close to maximal, which is excluded by the solar
neutrino oscillation data \cite{solnu}. This problem is circumvented in models
where neutrinos are induced at two loops \cite{Babu} or three loops \cite{KNT,
AKS, GNR}. One of the advantages of this class of models is that all the mass
scales are in the TeV or sub-TeV range, which makes it possible for them to be
tested at future colliders.

In Ref. \cite{KNT}, the SM was extended with two electrically charged
$SU(2)_{L}$ singlet scalars and one RH neutrino field, $N$, where a $Z_{2}$
symmetry was imposed to forbid the Dirac neutrino mass terms at tree level
\cite{KNT}. Once the electroweak symmetry is broken, neutrino masses are
generated at three loops, naturally explaining why their masses are so tiny
compared to the charged leptons as due to the high loop suppression. A
consequence of the $Z_{2}$ symmetry and the field content of the model, $N$ is
$Z_{2}$-odd, and thus guaranteed to be stable, which makes it a good DM
candidate. In Ref. \cite{Keung}, the authors considered extending the fermion
sector of the SM with two RH neutrinos, in order for it to be consistent with
the neutrino oscillation data, and they studied also its phenomenological implications.

Here, we calculate the three loop neutrino masses exactly, as compared to the
approximate expression derived in \cite{KNT}. We show that in order to satisfy
the recent experimental bound on the lepton flavor violating (LFV) process
such as $\mu\rightarrow e\gamma$ \cite{LFV}; and the anomalous magnetic moment
of the muon \cite{pdg}, one must have three generations of RH neutrinos.
Taking into account the neutrino oscillation data and the LFV constraints, we
show that the lightest RH neutrino can account for the DM abundance with
masses lighter than 225 \textrm{GeV}. The presence of the charged scalars in
this model will affect the Higgs decay process $h\rightarrow\gamma\gamma$ and
can lead to an enhancement with respect to the SM, where as $h\rightarrow
\gamma Z$ is slightly reduced. In this model, we find that a strongly
electroweak phase transition can be achieved with a Higgs mass of $\simeq125$
\textrm{GeV} as measured at the LHC \cite{ATLAS,CMS}.

This paper is organized as follows. In the next section we present the model,
and discuss the constraints from the LFV processes. In section III, we study
the relic density of the lightest RH neutrino, and discuss the coannihilation
effect due to the next lightest RH neutrino. The effect of the presence of
extra charged scalars on the Higgs decay channels $h\rightarrow\gamma\gamma
$\ and $h\rightarrow\gamma Z$ is discussed in section IV. Section V is devoted
to the study of the electroweak phase transition. In section VI, we discuss
the phenomenological implications of the RH neutrinos at electron-positron
colliders. Finally we conclude in section VII. The exact formula of the three
loop factor that enters in the expression of the neutrino masses is derived in
Appendix A. In Appendix B, we give the shift in masses for the gauge bosons
and the scalars at finite temperature.

\section{Neutrino Data and Flavor Violation Constraints}

In this section, we define the filed content of the model, give the exact
expression of the neutrino masses, and discuss the constraints from LFV processes.

\subsection{The Model}

Here we consider extending the SM with three RH neutrinos, $N_{i}$, and two
electrically charged scalars, $S_{1}^{\pm}$ and $S_{2}^{\pm}$, that are
singlet under $SU(2)_{L}$ gauge group. In addition, we impose a discrete
$Z_{2}$ symmetry on the model, under which $\{S_{2},N_{i}\}\rightarrow
\{-S_{2},-N_{i}\}$, and all other fields are even. The Lagrangian reads
\begin{align}
\mathcal{L}  &  =\mathcal{L}_{SM}+\{f_{\alpha\beta}L_{\alpha}^{T}Ci\tau
_{2}L_{\beta}S_{1}^{+}+g_{i\alpha}N_{i}S_{2}^{+}\ell_{\alpha R}\nonumber\\
&  +\tfrac{1}{2}m_{N_{i}}N_{i}^{C}N_{i}+h.c\}-V(\Phi,S_{1},S_{2}), \label{L}%
\end{align}
where $L_{\alpha}$ is the left-handed lepton doublet, $f_{\alpha\beta}$ are
Yukawa couplings which are antisymmetric in the generation indices $\alpha$
and $\beta$, $m_{N_{i}}$\ are the Majorana RH neutrino masses, $C$ is the
charge conjugation matrix, and $V(\Phi,S_{1},S_{2})$ is the tree-level scalar
potential which is given by
\begin{align}
V(\Phi,S_{1,2})  &  =\lambda\left(  \left\vert \Phi\right\vert ^{2}\right)
^{2}-\mu^{2}\left\vert \Phi\right\vert ^{2}+m_{1}^{2}S_{1}^{\ast}S_{1}%
+m_{2}^{2}S_{2}^{\ast}S_{2}+\lambda_{1}S_{1}^{\ast}S_{1}\left\vert
\Phi\right\vert ^{2}+\lambda_{2}S_{2}^{\ast}S_{2}\left\vert \Phi\right\vert
^{2}\nonumber\\
&  +\frac{\eta_{1}}{2}\left(  S_{1}^{\ast}S_{1}\right)  ^{2}+\frac{\eta_{2}%
}{2}\left(  S_{2}^{\ast}S_{2}\right)  ^{2}+\eta_{12}S_{1}^{\ast}S_{1}%
S_{2}^{\ast}S_{2}+\left\{  \lambda_{s}S_{1}S_{1}S_{2}^{\ast}S_{2}^{\ast
}+h.c\right\}  . \label{nu-mass}%
\end{align}
\qquad Here $\Phi$ denotes the SM Higgs doublet. It is worth mentioning that,
the charge breaking minima are not possible due to the positive-definite
values of $\lambda_{s}$ and $\eta_{12}$; in addition to the conditions on the
charged scalar masses $m_{S_{i}}^{2}=m_{i}^{2}+\lambda_{i}\upsilon^{2}/2>0$.

There are two immediate implications of the $Z_{2}$ symmetry imposed on the
Lagrangian:\newline

\begin{figure}[t]
\begin{centering}
\includegraphics[width=8cm,height=2.8cm]{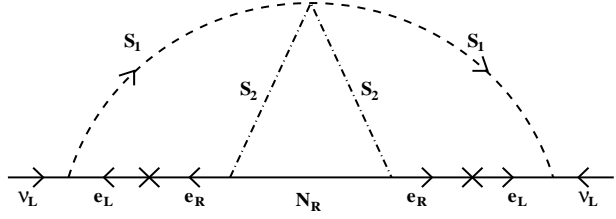}
\par\end{centering}
\caption{\textit{The three-loop diagram that generates the neutrino mass.}}%
\label{diag}%
\end{figure}

\begin{itemize}
\item First, if $N_{1}$ is the lightest particle among $N_{2}, N_{3}, S_{1}$
and $S_{2}$, then it would be stable, and hence it would be a candidate for
dark matter. Moreover, $N_{i}$ will be pair produced and subsequently decay
into $N_{1}$ (or to $N_{2}$ and then to $N_{1}$) and a pair (or two pairs) of
charged leptons. We will discuss its phenomenology in section VI.

\item The second implication, is that the Dirac neutrino mass term is
forbidden at all levels of the perturbation theory, and Majorana neutrinos
masses are generated radiatively at three-loops, as shown in Fig. \ref{diag}.
\end{itemize}

\subsection{Neutrino mass}

The neutrino mass matrix elements arising from the three-loop diagram in Fig.
\ref{diag}, are given by
\begin{equation}
(M_{\nu})_{\alpha\beta}=\frac{\lambda_{s}m_{\ell_{i}}m_{\ell_{k}}}{\left(
4\pi^{2}\right)  ^{3}m_{S_{2}}}f_{\alpha i}f_{\beta k}g_{ij}g_{kj}F\left(
m_{N_{j}}^{2}/m_{S_{2}}^{2},m_{S_{1}}^{2}/m_{S_{2}}^{2}\right)  ,
\label{nu-mass-1}%
\end{equation}
where $\rho,\kappa(=e,\mu,\tau)$ are the charged leptons flavor indices,
$i=1,2,3$ denotes the three right-handed neutrinos, and the function $F$ is a
loop integral given in (\ref{FF}), which was approximated to one in the
original work \cite{KNT}. Note that, unlike the conventional seesaw mechanism,
the radiatively generated neutrino masses are directly proportional to the
charged leptons and RH neutrino masses as shown in (\ref{nu-mass-1}) and
(\ref{FF}).

In general, the elements of the neutrino mass matrix can be written as%
\begin{equation}
(M_{\nu})_{\alpha\beta}=[U\cdot diag(m_{1},m_{2},m_{3})\cdot U^{T}%
]_{\alpha\beta}, \label{comp}%
\end{equation}
where $U$ is the Pontecorvo-Maki-Nakawaga-Sakata (PMNS) mixing matrix
\cite{PMNS}, which is parameterized in general by
\begin{equation}
U=\left(
\begin{array}
[c]{ccc}%
c_{12}c_{13} & c_{13}s_{12} & s_{13}e^{-i\delta_{D}}\\
-c_{23}s_{12}-c_{12}s_{13}s_{23}e^{i\delta_{D}} & c_{12}c_{23}-s_{12}%
s_{13}s_{23}e^{i\delta_{D}} & c_{13}s_{23}\\
s_{12}s_{23}-c_{12}c_{23}s_{13}e^{i\delta_{D}} & -c_{12}s_{23}-c_{23}%
s_{12}s_{13}e^{i\delta_{D}} & c_{13}c_{23}%
\end{array}
\right)  \allowbreak\left(
\begin{array}
[c]{ccc}%
1 & 0 & 0\\
0 & e^{i\alpha/2} & 0\\
0 & 0 & e^{i\beta/2}%
\end{array}
\right)  ,
\end{equation}
with $s_{ij}\equiv\sin(\theta_{ij})$ and $c_{ij}\equiv\cos(\theta_{ij}) $,
$\delta_{D}$ is the Dirac phase; and $\alpha$ and $\beta$ are the Majorana
phases. Using the experimental allowed values for $s_{12}^{2}=0.320_{-0.017}%
^{+0.016}$\textbf{, }$s_{23}^{2}=0.43_{-0.03}^{+0.03}$\textbf{, }$s_{13}%
^{2}=0.025_{-0.003}^{+0.003}$\textbf{, }$\left\vert \Delta m_{31}%
^{2}\right\vert =2.55_{-0.09}^{+0.06}\times10^{-3}$ \textrm{eV}$^{2}$ and
$\Delta m_{21}^{2}=7.62_{-0.19}^{+0.19}\times10^{-5}\mathrm{eV}^{2}$
\cite{GF}, we can find the parameter space of the model that is consistent
with the neutrino oscillation data.

\subsection{Experimental constraints}

Besides neutrino masses and mixing, the Lagrangian (\ref{L}) induces flavor
violating processes such as $\ell_{\alpha}\rightarrow\gamma\ell_{\beta}$ if
$m_{\ell_{\alpha}}>m_{\ell_{\beta}}$, generated at one loop via the exchange
of both extra charged scalars $S_{1,2}^{\pm}$. The branching ratio of such
process can be computed following \cite{MaRai} as \footnote{One has to mention
that this result is different from Eq. (38) in \cite{Keung}, where the authors
took the summation over the square of the $g_{i\alpha}$ terms instead of the
square of the their summation. The latter allows the parameter space of the
couplings to be enlarged.}
\begin{align}
B(\ell_{\alpha}  &  \rightarrow\gamma\ell_{\beta})=\frac{\Gamma(\ell_{\alpha
}\rightarrow\gamma\ell_{\beta})}{\Gamma(\ell_{\alpha}\rightarrow\ell_{\beta
}\nu_{\alpha}\bar{\nu}_{\beta})}\nonumber\\
&  =\frac{\alpha_{em}\upsilon^{4}}{384\pi}\left\{  \frac{\left\vert
f_{\kappa\alpha}f_{\kappa\beta}^{\ast}\right\vert ^{2}}{m_{S_{1}}^{4}}%
+\frac{36}{m_{S_{2}}^{4}}\left\vert {\sum\limits_{i}}g_{i\alpha}g_{i\beta
}^{\ast}F_{2}\left(  \frac{m_{N_{i}}^{2}}{m_{S_{2}}^{2}}\right)  \right\vert
^{2}\right\}  , \label{mutoegamma}%
\end{align}
with $\kappa\neq\alpha,\beta$, $\alpha_{em}$ is the fine structure constant
and $F_{2}(x)=(1-6x+3x^{2}+2x^{3}-6x^{2}\ln x)/6(1-x)^{4}$. For the case of
$\ell_{\alpha}=\ell_{\beta}=\mu$, this leads to a new contribution to the muon
anomalous magnetic moment $\delta a_{\mu}$, that is given by
\begin{equation}
\delta a_{\mu}=\frac{m_{\mu}^{2}}{16\pi^{2}}\left\{  \frac{\left\vert f_{\mu
e}\right\vert ^{2}+\left\vert f_{\mu\tau}\right\vert ^{2}}{6m_{S_{1}}^{2}%
}+\frac{1}{m_{S_{2}}^{2}}{\sum\limits_{i}}\left\vert g_{i\mu}\right\vert
^{2}F_{2}\left(  \frac{m_{N_{i}}^{2}}{m_{S_{2}}^{2}}\right)  \right\}  .
\end{equation}

\begin{figure}[t]
\begin{centering}
\includegraphics[width=8cm,height=6cm]{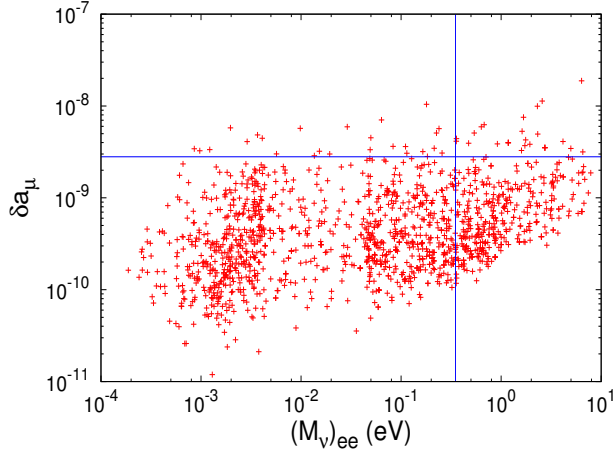}
\par\end{centering}
\caption{\textit{The muon anomalous magnetic moment versus the $\beta
\beta_{0\nu}$ decay effective Majorana }$\mathit{(M_{\nu})_{ee}}$\textit{. The
blue lines represent their experimental upper bounds.}}%
\label{amu}%
\end{figure}

In Fig. \ref{amu}, we show a scattered plot of the muon anomalous magnetic
moment versus the $\beta\beta_{0\nu}$ decay effective Majorana mass $\left(
M_{\nu}\right)  _{ee}$. In our scan of the parameter space of the model, we
took $m_{S_{1,2}}\geq100$ \textrm{GeV}\textbf{; }and demanded that
(\ref{nu-mass-1}) to be consistent with the neutrino oscillation data. From
Fig. \ref{amu}, one can see that most of the values of $\left(  M_{\nu
}\right)  _{ee}$ that are consistent with the bound on $\delta a_{\mu}$ are
lying in the range $10^{-3}$ \textrm{eV} to $\sim$\textrm{eV}. The current
bound on $\left(  M_{\nu}\right)  _{ee}$ is approximately $0.35~\mathrm{eV}$
\cite{bbo} and it is expected that within few years a number of next
generation $\beta\beta_{0\nu}$ experiments will be sensitive to $\left(
M_{\nu}\right)  _{ee}\sim10^{-2}$ \textrm{eV}\cite{futurebbo}.

\begin{figure}[t]
\begin{centering}
\includegraphics[width=8cm,height=6cm]{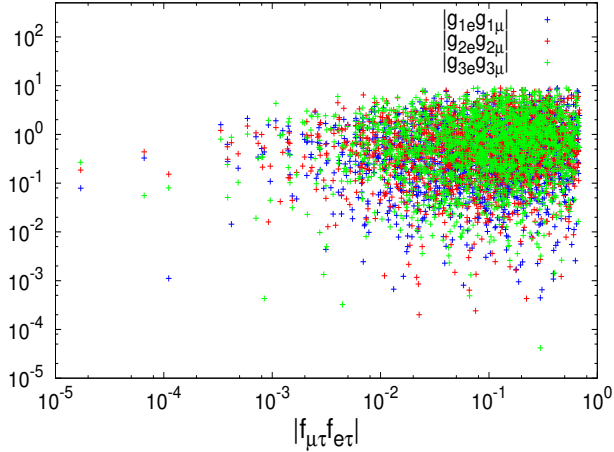}
\par\end{centering}
\caption{\textit{Different parameters combinations (as absolute values) that
are relevant to the LFV constrain on $B(\mu\rightarrow e\gamma)$, are shown
where (\ref{nu-mass-1}) and (\ref{comp}) are matched.}}%
\label{Par}%
\end{figure}

Fig. \ref{Par} gives an idea about the magnitude of the couplings that satisfy
the constraints from LFV processes and the muon anomalous magnetic moment, and
which also are consistent with the neutrino oscillation data. It is worth
noting that when considering just two generations of RH neutrinos (i.e,
$g_{3\alpha}=0$), we find that the bound $B\left(  \mu\rightarrow
e\gamma\right)  <5.7\times10^{-13}$ is violated \cite{LFV}\footnote{Although,
we have considered also the bound on $B(\tau\rightarrow\mu\gamma
)<4.5\times10^{-8}$ \cite{pdg}, but in our numerical scan, it does not
constrain severely the parameter space of the model.}. Therefore, having three
RH neutrinos is necessary for it to be in agreement with the data from the
bounds from LFV processes. Moreover, one has to mention that the bound on
$B\left(  \mu\rightarrow e\gamma\right)  $ makes the parameters space very
constrained. For instance, out of the benchmarks that are in agreement with
the neutrino oscillation data, DM and $\delta a_{\mu}$, only about 15\% of the
points will survive after imposing the $\mu\rightarrow e\gamma$ bound.

\section{Dark Matter, Coannihilation Effect \& Indirect Detection}

\subsection{Relic density}

As we noted earlier, the lightest RH neutrino $N_{1}$ is stable, and could be
the DM candidate. In the case of hierarchical RH neutrino mass spectrum, we
can safely neglect the effect of $N_{2}$ and $N_{3}$ on $N_{1}$ density. The
$N_{1}$ number density get depleted through the annihilation process
$N_{1}N_{1}\rightarrow\ell_{\alpha}\ell_{\beta}$ via the $t$-channel exchange
of $S_{2}^{\pm}$. For two incoming dark matter particles with momenta $p_{1}$
and $p_{2}$, and final states charged leptons with momenta $k_{1}$ and $k_{2}%
$, the amplitude for this process is
\begin{equation}
\mathcal{M}_{\alpha\beta}=g_{1\alpha}g_{1\beta}^{\ast}\left[  \frac{{\bar{u}%
}(k_{1})P_{L}u(p_{1}).\bar{v}(p_{2})P_{R}v(k_{2})}{t-m_{S_{2}}^{2}}%
-\frac{{\bar{u}}(k_{1})P_{L}u(p_{2}).\bar{v}(p_{1})P_{R}v(k_{2})}{u-m_{S_{2}%
}^{2}}\right]  ,
\end{equation}
where $t=(p_{1}-k_{1})^{2}$ and $u=(p_{1}-k_{2})^{2}$ are the Mandelstam
variables corresponding the t and u channels, respectively. After squaring,
summing and averaging over the spin states, we find that in the
non-relativistic limit, the total annihilation cross section is given by
\begin{equation}
\sigma_{N_{1}N_{1}}\upsilon_{r}\simeq\sum_{\alpha,\beta}|g_{1\alpha}g_{1\beta
}^{\ast}|^{2}\frac{m_{N_{1}}^{2}\left(  m_{S_{2}}^{4}+m_{N_{1}}^{4}\right)
}{48\pi\left(  m_{S_{2}}^{2}+m_{N_{1}}^{2}\right)  ^{4}}\upsilon_{r}^{2},
\label{sig11}%
\end{equation}
with $\upsilon_{r}$ is the relative velocity between the annihilation $N_{1}%
$'s. As the temperature of the universe drops below the freeze-out temperature
$T_{f}\sim m_{N_{1}}/{25}$, the annihilation rate becomes smaller than the
expansion rate (the Hubble parameter) of the universe, and the $N_{1}$'s start
to decouple from the thermal bath. The relic density after the decoupling can
be obtained by solving the Boltzmann equation, and it is approximately given
by
\begin{align}
\Omega_{N_{1}}h^{2}  &  \simeq\frac{2x_{f}\times1.1\times10^{9}\mathrm{GeV}%
^{-1}}{\sqrt{g_{\ast}}M_{pl}\left\langle \sigma_{N_{1}N_{1}}\upsilon
_{r}\right\rangle }\label{Omh2}\\
&  \simeq\frac{1.28\times10^{-2}}{\sum_{\alpha,\beta}|g_{1\alpha}g_{1\beta
}^{\ast}|^{2}}\left(  \frac{m_{N_{1}}}{135~\mathrm{GeV}}\right)  ^{2}%
\frac{\left(  1+m_{S_{2}}^{2}/m_{N_{1}}^{2}\right)  ^{4}}{1+m_{S_{2}}%
^{4}/m_{N_{1}}^{4}},\nonumber
\end{align}
where ${<\upsilon_{r}^{2}>}\simeq6/x_{f}\simeq6/25$ is the thermal average of
the relative velocity squared of a pair of two $N_{1}$ particles, $M_{pl}$ is
planck mass; and $g_{\ast}(T_{f})$ is the total number of effective massless
degrees of freedom at $T_{f}$\textbf{.}

\begin{figure}[t]
\label{Om}\begin{centering}
\includegraphics[width=8cm,height=6cm]{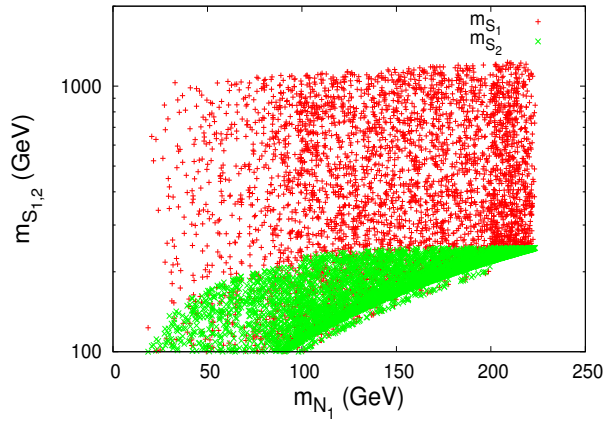}
\par\end{centering}
\caption{\textit{The charged scalar masses $m_{S_{1}}$ (red) and $m_{S_{2}}$
(green) versus the lightest RH neutrino mass, where the consistency with the
neutrino data, LFV constraints and the DM relic density have been imposed.}}%
\label{msmn}%
\end{figure}

In Fig. \ref{msmn}, we plot the allowed mass range $(m_{N_{1}},m_{S_{i}})$
plane that give the observed dark matter relic density \cite{Planck}. As seen
in the figure, the neutrino experimental data combined with the relic density
seems to prefer $m_{S_{1}}>m_{S_{2}}$ for large space of parameters. However,
the masses of both the DM and the charged scalar $S_{2}^{\pm}$ can not exceed
$m_{N_{1}}<225$ \textrm{GeV}\ and $m_{S_{2}}<245$ \textrm{GeV},
respectively.\textbf{\ }

\subsection{Coannihilation effect}

In computing the relic density in (\ref{Omh2}), we have assumed that there is
a hierarchy between the three right-handed neutrino masses. However, if we
consider the possibility for $N_{2}$ and/or $N_{3}$ being close in mass to
$N_{1}$, i.e $\Delta_{i}=(m_{N_{i}}-m_{N_{1}})/m_{N_{1}}<<1$, then
coannihilation processes like $N_{1}N_{2,3}\rightarrow\ell_{\alpha}\ell
_{\beta}$ might have important effects on the evolution of the $N_{1}$-number
density. The process $N_{1}S_{2}^{\pm}\rightarrow\ell_{\alpha}^{\pm}\gamma$ is
suppressed by the large mass difference between $S_{2}^{\pm}$ and $N_{1}$ and
the smallness of the electromagnetic coupling compared to $\mathcal{O}(g^{2}%
)$, and therefore its contribution to the coannihilation is negligible.

Following \cite{Griest}, the coannihilation effect could be included by
re-evaluating the relic density (\ref{Omh2}) using the effective annihilation
cross section and multiplicity
\begin{equation}%
\begin{array}
[c]{c}%
\sigma_{eff}(x)={\sum_{i,k}^{3}}\tfrac{2^{2}}{g_{eff}^{2}}\left(  1+\Delta
_{i}\right)  ^{3/2}\left(  1+\Delta_{k}\right)  ^{3/2}e^{-x\left(  \Delta
_{i}+\Delta_{k}\right)  }\sigma\left(  N_{i}N_{k}\rightarrow\ell_{\alpha}%
^{-}\ell_{\beta}^{+}\right)  ,\\
g_{eff}(x)={\sum_{i}^{3}}2\left(  1+\Delta_{i}\right)  ^{3/2}e^{-x\Delta_{i}},
\end{array}
\label{sigeff}%
\end{equation}
For freeze-out temperature much smaller than $m_{N_{1}}$, the effective cross
section can be written as $\sigma_{eff}(x)\upsilon_{r}=a_{eff}(x)+b_{eff}%
(x)\upsilon_{r}^{2}+\mathcal{O}\left(  \upsilon_{r}^{4}\right)  $, where
\begin{align}
a_{eff}(z)  &  ={\sum_{ik}}a_{ik}\tfrac{2^{2}}{g_{eff}^{2}(z)}\left(
1+\Delta_{i}\right)  ^{3/2}\left(  1+\Delta_{k}\right)  ^{3/2}e^{-z\left(
\Delta_{i}+\Delta_{k}\right)  },\nonumber\\
b_{eff}(z)  &  ={\sum_{ik}}b_{ik}\tfrac{2^{2}}{g_{eff}^{2}(z)}\left(
1+\Delta_{i}\right)  ^{3/2}\left(  1+\Delta_{k}\right)  ^{3/2}e^{-z\left(
\Delta_{i}+\Delta_{k}\right)  }.
\end{align}
Here the factors $a_{ik}$\ and $b_{ik}$\ correspond to the two first terms in
the velocity expansion of $\sigma\left(  N_{i}N_{k}\rightarrow\ell_{\alpha
}^{+}\ell_{\beta}^{-}\right)  \upsilon_{r}$ (i.e, the $s$ and $p$ wave terms,
respectively), given by%
\begin{align}
a_{ik}  &  =\tfrac{1}{32\pi}{\sum\limits_{\alpha,\beta}}\left\vert g_{i\alpha
}g_{k\beta}^{\ast}-g_{i\beta}^{\ast}g_{k\alpha}\right\vert ^{2}\frac{m_{N_{i}%
}m_{N_{k}}}{\left(  m_{N_{i}}^{2}+m_{S_{2}}^{2}\right)  \left(  m_{N_{k}}%
^{2}+m_{S_{2}}^{2}\right)  },\nonumber\\
b_{ik}  &  =\frac{m_{N_{i}}m_{N_{k}}}{48\pi\left(  m_{N_{i}}^{2}+m_{S_{2}}%
^{2}\right)  ^{2}\left(  m_{N_{k}}^{2}+m_{S_{2}}^{2}\right)  ^{2}}\left\{
{\sum\limits_{\alpha,\beta}}\left\vert g_{i\alpha}g_{k\beta}^{\ast}\right\vert
^{2}\left(  m_{N_{i}}^{2}m_{N_{k}}^{2}+m_{S_{2}}^{4}\right)  \right.
\nonumber\\
&  \left.  +\tfrac{1}{2}{\sum\limits_{\alpha,\beta}}\left\vert g_{i\alpha
}g_{k\beta}^{\ast}-g_{i\beta}^{\ast}g_{k\alpha}\right\vert ^{2}\left(
m_{S_{2}}^{4}-3m_{N_{i}}m_{N_{k}}m_{S_{2}}^{2}-m_{N_{i}}^{2}m_{N_{k}}%
^{2}\right)  \right\}  .
\end{align}
Thus, the coannihilation effect on the relic density could be accounted for by
just multiplying the couplings term $\sum_{\alpha,\beta}|g_{1\alpha}g_{1\beta
}^{\ast}|^{2}$\ in (\ref{Omh2}) by the factor
\begin{equation}
\kappa=\left(  \frac{x_{f}^{\prime}}{x_{f}}\right)  ^{-1}\left(  \frac
{g_{\ast}(x_{f}^{\prime})}{g_{\ast}(x_{f})}\right)  ^{1/2}\frac{I_{a}%
(x_{f}^{\prime})+I_{b}(x_{f}^{\prime})\upsilon_{r}^{\prime2}}{b_{11}%
\upsilon_{r}^{2}}. \label{kapa}%
\end{equation}
Here $x_{f}^{\prime}=m_{N_{1}}/T_{f}^{\prime}$ is the freeze-out temperature
defined using $\sigma_{eff}(x)\upsilon_{r}$ instead of $\sigma_{N_{1}N_{1}%
}\upsilon_{r}$, and the integral functions $I_{a}\left(  x\right)  $\ and
$I_{b}\left(  x\right)  $ are given by
\begin{equation}
I_{a}\left(  x\right)  =x{\int_{x}^{\infty}}a_{eff}(z)z^{-2}dz,~I_{b}\left(
x\right)  =2x^{2}{\int_{x}^{\infty}}b_{eff}(z)z^{-3}dz,
\end{equation}
which, in general, causes a shift in the freeze-out temperature. In our case,
we find that $x_{f}$ gets lowered by less than $3\%$ for $5\%<\Delta
_{2,3}<25\%$, and therefore, within this range, the effect of coannihilation
on the freeze-out temperature and $g_{\ast}(x_{f})$ can be ignored and one
approximates the factor $\kappa$ by the ratio $(I_{a}(x_{f})+I_{b}%
(x_{f})\upsilon_{r}^{2})/b_{11}\upsilon_{r}^{2}$. However for $\Delta
_{2,3}<<5\%$, $x_{f}$ can be shifted by more than a factor of two, which
results in a freeze-out temperature of about $50~\mathrm{MeV}$. Hence, for
$N_{1}$ to have the observed cosmological relic density, it must be that
\begin{equation}
\kappa\sum_{\alpha,\beta}|g_{1\alpha}g_{1\beta}|^{2}=(114.04\pm3.56\,)\times
10^{-3}\times\left(  \frac{m_{N_{1}}}{135~\mathrm{GeV}}\right)  ^{2}%
\frac{\left(  1+m_{S_{2}}^{2}/m_{N_{1}}^{2}\right)  ^{4}}{1+m_{S_{2}}%
^{4}/m_{N_{1}}^{4}}, \label{Omega}%
\end{equation}

\begin{figure}[t]
\begin{centering}
\includegraphics[width=8cm,height=6cm]{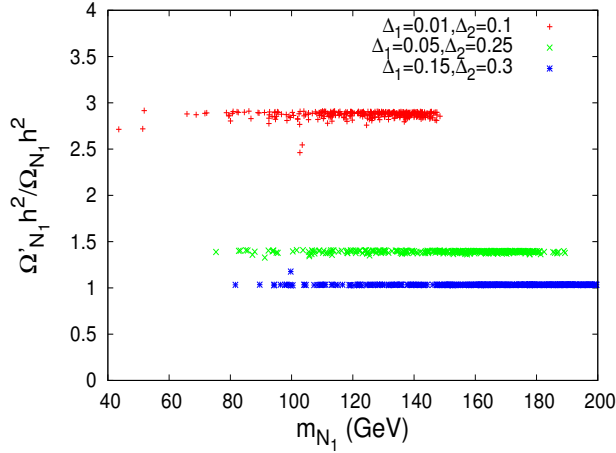}
\par\end{centering}
\caption{\textit{The ratio }$\Omega_{N_{1}}^{\prime}h^{2}/\Omega_{N_{1}}h^{2}%
$\textit{\ versus the lightest RH neutrino mass }$m_{N_{1}}$\textit{, where
}$\Omega_{N_{1}}h^{2}$\textit{\ (}$\Omega_{N_{1}}^{\prime}h^{2}$\textit{) is
the }$N_{1}$\textit{\ relic density without (with) the coannihilation
effect.}}%
\label{kp}%
\end{figure}

In Fig. \ref{kp}, we plot the ratio $\Omega_{N_{1}}^{\prime}h^{2}%
/\Omega_{N_{1}}h^{2}$ versus $m_{N_{1}}$, where $\Omega_{N_{1}}h^{2}%
(\Omega_{N_{1}}^{\prime}h^{2})$ is the relic density estimated without (with)
coannihilation effect. We see that for $\Delta_{1}\lesssim0.05$, the
coannihilation becomes significant and leads to an increase in the relic
density by more than $50\%$, whereas for $\Delta_{1}\lesssim0.01$, this
increase is almost factor of three. 

\subsection{Indirect Detection constrains}

Before closing this section, we would like to comment on the detection of the
dark matter in our model.

\begin{itemize}
\item The direct detection: Since the interactions of $N_{1}$ involve only
leptons, the $N_{1}$-nucleon scattering is absent at the tree level. Moreover,
it can not scatter via electromagnetic interaction since the dipole moment for
Majorana particles vanish identically. However, if $N_{1}$ and $N_{2}$ (or
$N_{3}$) are quasi-degenerate, then a transition magnetic dipole moment can be
generated radiatively, and in that case an inelastic scattering $N_{1}%
+p\rightarrow N_{2}+p$ is possible, provided that $\Delta_{1}\leq
10^{-6}\left(  \frac{10~\mathrm{GeV}}{m_{N_{1}}}\right)  $. It is quite
unlikely that such a tiny degeneracy will be stable under the radiative
correction to $m_{N_{1}}$ and $m_{N_{2}}$ generated via one loop diagrams by
the exchange of $S_{2}$ and charged leptons.

\item The indirect detection: The $N_{1}$ annihilation rate into leptons has a
helicity suppressed $s$-wave term (i.e. $\propto m_{l}^{2}/m_{N_{1}}^{2}$) and
a p-wave contribution. For $m_{N_{1}}\geq100$ \textrm{GeV}, both terms can be
of the same order for $g_{1\tau}\sim1$. Unfortunately, with the dark matter
velocity in the galactic halo of the order $10^{-3}$, the cross section is too
small to have a chance for the annihilation products of $N_{1}$ to be
detected. This helicity suppression of the $s$-wave will be lifted if the
final state has in addition a spin one particle. This is the case for the
internal bremsstrahlung (IB) processes, where a photon is emitted from the
final state charged leptons (FSR) or from charged mediator in the $t$-channel
propagator (VIB ) \footnote{Although the soft and collinear FSR is
logarithmically enhanced, it is helicity suppressed, and thus typically
smaller than VIB . However, they have to be both included to obtain a gauge
invariant result.}. The later exhibits an enhancement for large photon
energies if the dark matter and the particle in the propagator are almost
degenerate in mass, whereas the FSR is dominated by the photons that are
approximately collinear with either $\ell_{\alpha}$ or $\ell_{\beta}$. The
annihilation process $N_{1}N_{1}\rightarrow\ell_{\alpha}\ell_{\beta}\gamma$
not only could have a larger cross section than $N_{1}N_{1}\rightarrow
\ell_{\alpha}\ell_{\beta}$ but also leads to a gamma ray signal with sharp
spectral features that are potentially observable at future gamma-ray
telescopes. Very recently \cite{Berg0}, it has been shown that in this model,
a DM with the mass $m_{N_{1}}\sim$135 $\mathrm{GeV}$ could lead to the
following three effects: (1) a wide internal bremsstrahlung bump with maximum
of $90\%m_{N_{1}}\sim$120 $\mathrm{GeV}$, (2) a $\gamma\gamma$ line around
$E_{\gamma}=m_{N_{1}}\sim$135 $\mathrm{GeV}$, and (3) a $Z\gamma$ line at
$E_{\gamma}=m_{N_{1}}(1-m_{Z}^{2}/(4m_{N_{1}}^{2}))\sim$119.6 $\mathrm{GeV}$.
These features together provide a good fit to the gamma rays excess observed
in the Fermi-LAT data \cite{Berg0}.
\end{itemize}

\section{The Higgs decay channels $h\rightarrow\gamma\gamma$ and
$h\rightarrow\gamma Z$}

Recently, ATLAS \cite{ATLAS2} and CMS \cite{CMS2} collaborations have
announced the observation of a scalar particle with mass $\simeq125$
\textrm{GeV} at about $5~\sigma$ confidence level. The question is whether or
not this is really the SM Higgs or some Higgs-like state with different
properties. Indeed, the fit of the data by the ATLAS collaboration seems to
show an excess in $h\rightarrow\gamma\gamma$ events by more than $50\%$ with
respect to the SM, while the updated CMS analysis is consistent with the SM.
Defining $R_{\gamma\gamma}$ to be the decay width of $h\rightarrow\gamma
\gamma$ scaled by its expected SM value, we find that
\begin{equation}
R_{\gamma\gamma}=\left\vert 1+\frac{\upsilon^{2}}{2}\frac{\frac{\lambda_{1}%
}{m_{S_{1}}^{2}}A_{0}\left(  \tau_{S_{1}}\right)  +\frac{\lambda_{2}}%
{m_{S_{2}}^{2}}A_{0}\left(  \tau_{S_{2}}\right)  }{A_{1}\left(  \tau
_{W}\right)  +N_{c}Q_{t}^{2}A_{1/2}\left(  \tau_{t}\right)  }\right\vert ^{2},
\label{Ryy}%
\end{equation}
where $\tau_{X}=m_{h}^{2}/4m_{X}^{2}$, with $m_{X}$ is the mass of the charged
particle X running in the loop, $N_{c}=3$ is the color number, and $Q_{t}$ is
the electric charge of the top quark in unit of $\left\vert e\right\vert $.
The loop amplitudes $A_{i}$ for spin $0$, spin $1/2$ and spin $1$ particle
contribution are given by \cite{djouadi}%
\begin{align}
A_{0}\left(  x\right)   &  =-x^{-2}\left[  x-f\left(  x\right)  \right]
,\nonumber\label{A}\\
A_{1/2}\left(  x\right)   &  =2x^{-2}\left[  x+\left(  x-1\right)  f\left(
x\right)  \right]  ,\nonumber\\
A_{1}\left(  x\right)   &  =-x^{-2}\left[  2x^{2}+3x+3\left(  2x-1\right)
f\left(  x\right)  \right]  ,
\end{align}
with%
\begin{equation}
f\left(  x\right)  =\left\{
\begin{array}
[c]{ccc}%
\arcsin^{2}\left(  \sqrt{x}\right)  &  & x\leq1\\
-\frac{1}{4}\left[  \log\frac{1+\sqrt{1-x^{-1}}}{1-\sqrt{1-x^{-1}}}%
-i\pi\right]  ^{2} &  & x>1.
\end{array}
\right.
\end{equation}
The effect on the $B(h\rightarrow\gamma\gamma)$ charged scalar singlets will
depend on how light are $S_{1,2}^{\pm}$, the sign and the strength of their
couplings to the SM Higgs doublet. For instance, an enhancement can be
achieved by taking $\lambda_{1}$ and/or $\lambda_{2}$ to be negative.

Another Higgs decay channel that could be modified due to these extra charged
fields, is $h\rightarrow\gamma Z$, where similarly the effect is parameterized
by%
\begin{equation}
R_{\gamma Z}=\left\vert 1+s_{w}^{2}\frac{\upsilon^{2}}{2}\frac{\frac
{\lambda_{1}}{m_{S_{1}}^{2}}A_{0}\left(  \tau_{S_{1}},\zeta_{S_{1}}\right)
+\frac{\lambda_{2}}{m_{S_{2}}^{2}}A_{0}\left(  \tau_{S_{2}},\zeta_{S_{2}%
}\right)  }{c_{w}A_{1}\left(  \tau_{W},\zeta_{W}\right)  +\frac{2\left(
1-8s_{w}^{2}/3\right)  }{c_{w}}A_{1/2}\left(  \tau_{t},\zeta_{t}\right)
}\right\vert ^{2}, \label{RyZ}%
\end{equation}
where $\zeta_{X}=m_{Z}^{2}/4m_{X}^{2}$, and the $A$'s functions here are given
by \cite{djouadi}%
\begin{align}
A_{0}\left(  x,y\right)   &  =I_{1}\left(  x,y\right)  ,\nonumber\\
A_{1/2}\left(  x,y\right)   &  =I_{1}\left(  x,y\right)  -I_{2}\left(
x,y\right)  ,\nonumber\\
A_{1}\left(  x,y\right)   &  =\left[  \left(  1+2x\right)  \tan^{2}\theta
_{w}-\left(  5+2x\right)  \right]  I_{1}\left(  x,y\right)  +4\left(
3-\tan^{2}\theta_{w}\right)  I_{2}\left(  x,y\right)  ,
\end{align}
with
\begin{align}
I_{1}\left(  x,y\right)   &  =-\tfrac{1}{2\left(  x-y\right)  }+\tfrac
{f\left(  x\right)  -f\left(  y\right)  }{2\left(  x-y\right)  ^{2}}%
+\tfrac{y\left[  g\left(  x\right)  -g\left(  y\right)  \right]  }{\left(
x-y\right)  ^{2}},\nonumber\\
I_{2}\left(  x,y\right)   &  =\tfrac{f\left(  x\right)  -f\left(  y\right)
}{2\left(  x-y\right)  },
\end{align}
and
\begin{equation}
g\left(  x\right)  =\left\{
\begin{array}
[c]{ccc}%
\sqrt{x^{-1}-1}\arcsin\left(  \sqrt{x}\right)  &  & x\leq1\\
\frac{\sqrt{1-x^{-1}}}{2}\left[  \log\frac{1+\sqrt{1-x^{-1}}}{1-\sqrt
{1-x^{-1}}}-i\pi\right]  &  & x>1.
\end{array}
\right.
\end{equation}

\begin{figure}[t]
\begin{centering}
\includegraphics[width=8cm,height=6cm]{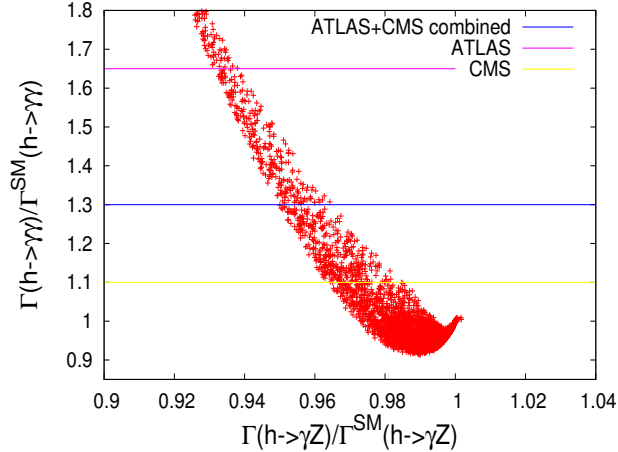}
\par\end{centering}
\caption{\textit{The modified Higgs decay rates $B(h\rightarrow\gamma\gamma)$
vs $B(h\rightarrow\gamma Z)$, scaled by their SM values, due to the extra
charged scalars, for randomly chosen sets of parameters. The magenta (yellow)
line represents the ATLAS (CMS) recent measurements on the $h\rightarrow
\gamma\gamma$ channel, while the blue one is their combined result.}}%
\label{Hr}%
\end{figure}

In Fig. \ref{Hr}, we present $R_{\gamma\gamma}$ versus $R_{\gamma Z}$ for
randomly chosen sets of parameters where the charged scalars are taken to be
heavier than 100 $\mathrm{GeV}$, the Higgs mass within the range
$124<m_{h}<126~\mathrm{GeV}$, and the condition of a strongly first order
phase transition is implemented (see next section). In our numerical scan, we
take the model parameters relevant for the Higgs decay to be in the range%
\begin{equation}
\lambda<2,~\left\vert \lambda_{1,2}\right\vert <3,~m_{1,2}^{2}<2~\mathrm{TeV}%
^{2}, \label{par}%
\end{equation}
where the Higgs mass is calculated at one-loop level. An enhancement of
$B(h\rightarrow\gamma\gamma)$ can be obtained for a large range of parameter
space, whereas $B(h\rightarrow\gamma Z)$ is slightly reduced with respect to
the SM. It is interesting to note that if one consider the combined ATLAS and
CMS di-photon excess, then $R_{\gamma Z}$ is predicted to be smaller than the
expected SM value by approximately $5\%$.

\section{A Strong First Order Electroweak Phase Transition}

It is well known that the SM has all the qualitative ingredients for
electroweak baryogenesis, but the amount of matter-antimatter asymmetry
generated is too small. One of the reasons is that the electroweak phase
transition (EWPT) is not strongly first order, which is required to suppress
the sphaleron processes in the broken phase. The strength of the EWPT can be
improved if there are new scalar degrees of freedom around the electroweak
scale coupled to the SM Higgs, which is the case in the model that we are
considering in this paper.

The investigation of the transition dynamics and its strength requires the
precise knowledge of the effective potential of the CP-even scalar fields at
finite temperature \cite{Th}. The zero temperature one-loop Higgs effective
potential is given in the $\overline{DR}$ scheme by
\begin{equation}
V^{T=0}(h)=\frac{\lambda}{4!}h^{4}-\frac{\mu^{2}}{2}h^{2}+\sum_{i}n_{i}%
\frac{m_{i}^{4}(h)}{64\pi^{2}}\left(  \ln\left(  \frac{m_{i}^{2}(h)}%
{\Lambda^{2}}\right)  -\frac{3}{2}\right)  ,
\end{equation}
where $h=(\sqrt{2}Re(H^{0})-\upsilon)$\ is the real part of the neutral
component in the doublet, $n_{i}$\ are the field multiplicities, $m_{i}%
^{2}(h)$ are the field-dependent mass squared which are given in Appendix B,
and $\Lambda$ is the renormalization scale which we choose to be the top quark
mass.\ At tree-level, the parameter $\mu^{2}$ in the potential is given by
$\mu^{2}=\lambda\upsilon^{2}$, but if the one-loop corrections are considered,
the parameter $\mu^{2}$ is corrected by the counter-term
\begin{equation}
\delta\mu^{2}=\sum_{i}\frac{n_{i}}{\upsilon}\left.  \frac{dm_{i}^{2}}%
{d\tilde{h}}\frac{m_{i}^{2}}{32\pi^{2}}\left(  \ln\left(  \frac{m_{i}^{2}%
}{\Lambda^{2}}\right)  -1\right)  \right\vert _{h=\upsilon,\mu^{2}\equiv
\mu^{2}+\delta\mu^{2}},
\end{equation}

For instance, the one loop correction to the Higgs mass due to the charged
singlets, when neglecting the Higgs and gauge bosons contributions, is
\begin{equation}
m_{h}^{2}\simeq2\lambda\upsilon^{2}+{\sum\limits_{i}}\frac{\lambda_{i}%
^{2}\upsilon^{2}}{16\pi^{2}}\ln\frac{m_{S_{i}}^{2}}{m_{t}^{2}}, \label{mh}%
\end{equation}
where the first term on the right hand side of the equation is the Higgs mass
at the tree level. If one takes $m_{S_{1}}=m_{S_{2}}=2m_{t}$ and $\lambda
_{1}=\lambda_{2}$, then the Higgs mass is exactly 125$\;\mathrm{GeV}$ for
$\lambda=10^{-1}$, $10^{-2}$, $10^{-3}$ if $\lambda_{1}=1.82$, $3.68$, $3.82$,
respectively. Note that these values are still within the perturbative regime.
On the other hand, these extra corrections could be negative and may relax the
large tree-level mass value of the Higgs to its experimental value for
$\lambda$ large. Therefore, it is expected that these extra charged scalars
will help the EWPT to be strongly first order by enhancing the value of the
effective potential at the wrong vacuum at the critical temperature without
suppressing the ratio $\upsilon(T_{c})/T_{c}$, and therefore avoiding the
severe bound on the mass of the SM Higgs. However, as it has been shown in
section $3.1$, the relic density requires large values for $m_{S_{1}}$ and so
the Higgs mass in Eq (\ref{mh}) can be easily set to its experimental value
($125~GeV$),while keeping $S_{2}$ light, for small doublet quartic coupling
(which gives a strong EWPT). Thus, both the measured values of the Higgs mass
and the requirement for the EWPT to be strongly first order are not in
conflict with values of $m_{2}$ smaller than $245~GeV$ (as required from the
observed relic density).

In order to generate a baryon asymmetry at the electroweak scale \cite{EWB},
the anomalous violating $B+L$ interactions should be switched-off inside the
nucleated bubbles, which implies the famous condition for a strong first order
phase transition \cite{SFOPT}
\begin{equation}
\upsilon(T_{c})/T_{c}>1, \label{v/t}%
\end{equation}
where $T_{c}$ is the critical temperature at which the effective potential
exhibits two degenerate minima, one at zero and the other at $\upsilon(T_{c})
$. Both $T_{c}$ and $\upsilon(T_{c})$ are determined using the full effective
potential at finite temperature \cite{Th}
\begin{equation}
V_{eff}(h,T)=V^{T=0}(h)+\tfrac{T^{4}}{2\pi^{2}}\sum_{i}n_{i}J_{B,F}\left(
m_{i}^{2}/T^{2}\right)  +V_{ring}(h,T);
\end{equation}
with
\begin{equation}
J_{B,F}\left(  \alpha\right)  =\int_{0}^{\infty}x^{2}\log(1\mp\exp
(-\sqrt{x^{2}+\alpha})), \label{JBF}%
\end{equation}
and%
\begin{equation}
V_{ring}(h,T)=-\frac{T}{12\pi}\sum\limits_{i}n_{i}\left\{  \tilde{m}_{i}%
^{3}(h,T)-m_{i}^{3}(h)\right\}  , \label{ring}%
\end{equation}
where the summation is performed over the scalar longitudinal gauge degrees of
freedom, and $\tilde{m}_{i}^{2}(h,T)$ are their thermal masses, which are
given in Appendix B. The contribution (\ref{ring}) is obtained by performing
the resummation of an infinite class of of infrared divergent multi-loops,
known as the ring (or daisy) diagrams, which describes a dominant contribution
of long distances and gives significant contribution when massless states
appear in a system. It amounts to shifting the longitudinal gauge boson and
the scalar masses obtained by considering only the first two terms in the
effective potential \cite{ring}. This shift in the thermal masses of
longitudinal gauge bosons and not their transverse parts tends to reduce the
strength of the phase transition. The integrals (\ref{JBF}) is often estimated
in the high temperature approximation, however, in order to take into account
the effect of all the (heavy and light) degrees of freedom, we evaluate them numerically.

In the SM, the ratio $\upsilon(T_{c})/T_{c}$ is approximately $\left(
2m_{W}^{3}+m_{Z}^{3}\right)  /\left(  \pi\upsilon m_{h}^{2}\right)  $, and
therefore the criterion for a strongly first phase transition is not fulfilled
for $m_{h}>42~\mathrm{GeV}$. However, if the one-loop corrections in
(\ref{mh}) are sizeable, then this bound could be relaxed in such a way that
the Higgs mass is consistent with the measured value at the LHC. This might be
possible since the extra charged scalars affect the dynamics of the SM scalar
field vev around the critical temperature \cite{hna}. \begin{figure}[t]
\begin{center}
\includegraphics[width=8cm,height=6cm]{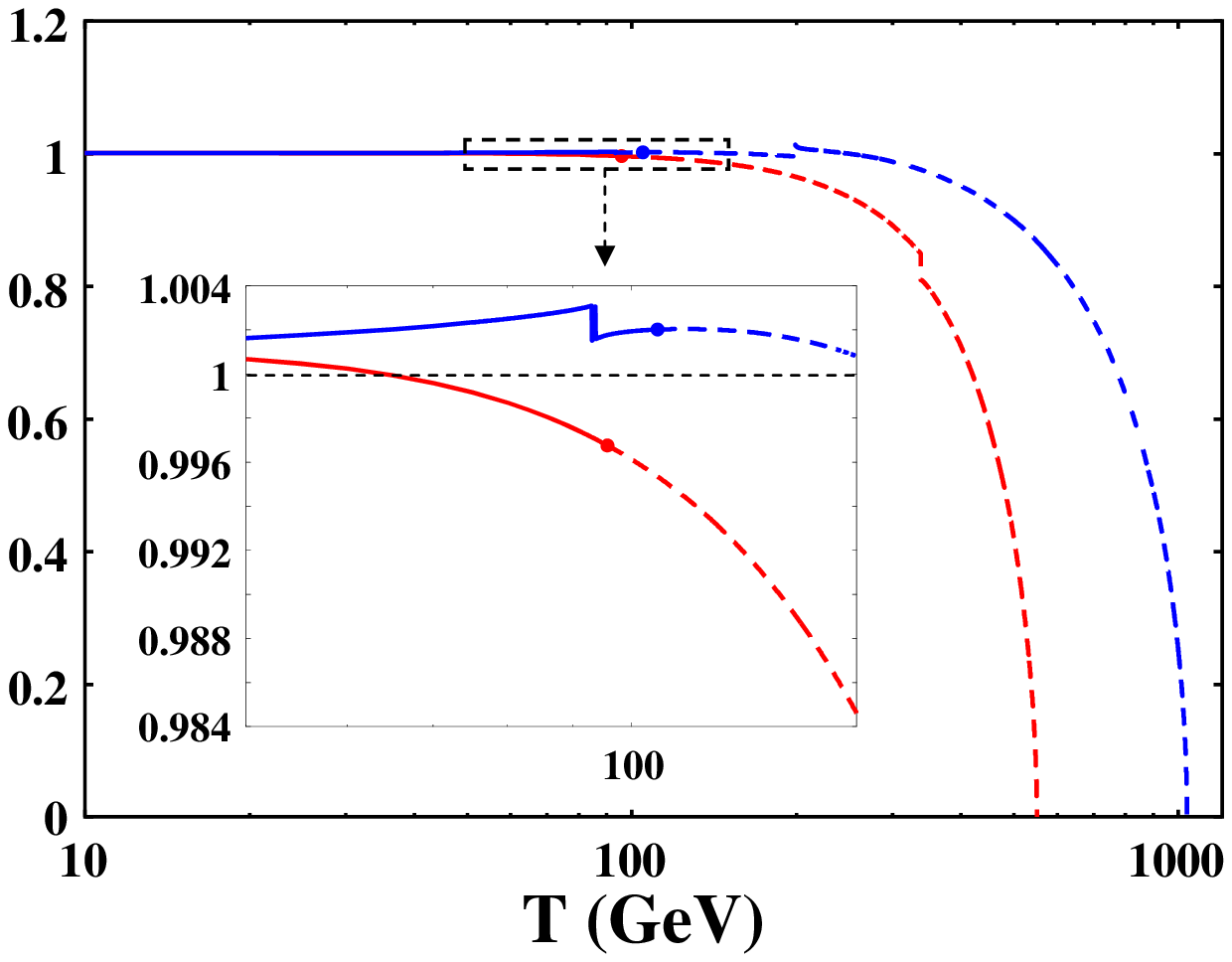}
\end{center}
\caption{\textit{The dependance of the Higgs vev scaled by the zero
temperature value $\upsilon=246$ GeV, on the temperature below (solid lines)
and above (dashed lines) the critical temperature for two benchmarks, where
the red (blue) one corresponds to small (large) $\lambda$ value and the
positive (negative) scalar contributions in (\ref{mh}) relax the Higgs mass to
its experimental value.}}%
\label{vac}%
\end{figure}

This is shown in Fig. \ref{vac}, where one sees the evolution of $\upsilon(T)
$ with respect to the temperature. In contrast to the SM, where the EW vev
decays quickly to zero just around $T\sim100\;\mathrm{GeV}$, here it is
delayed up to $\mathrm{TeV}$ due to the existence of the extra charged
scalars. This can be understood due to the fact that the value of the
effective potential at the wrong vacuum ( $<h>=0$) is temperature-dependant
through the charged scalars thermal masses in the symmetric phase. The
evolution of the effective potential at this (wrong) minimum makes the
transition happening at $T\geq100\;\mathrm{GeV}$, while the Higgs vev is
slowly decaying with respect to the temperature as shown in Fig.\ref{vac}.

\begin{figure}[t]
\includegraphics[width=8cm,height=6cm]{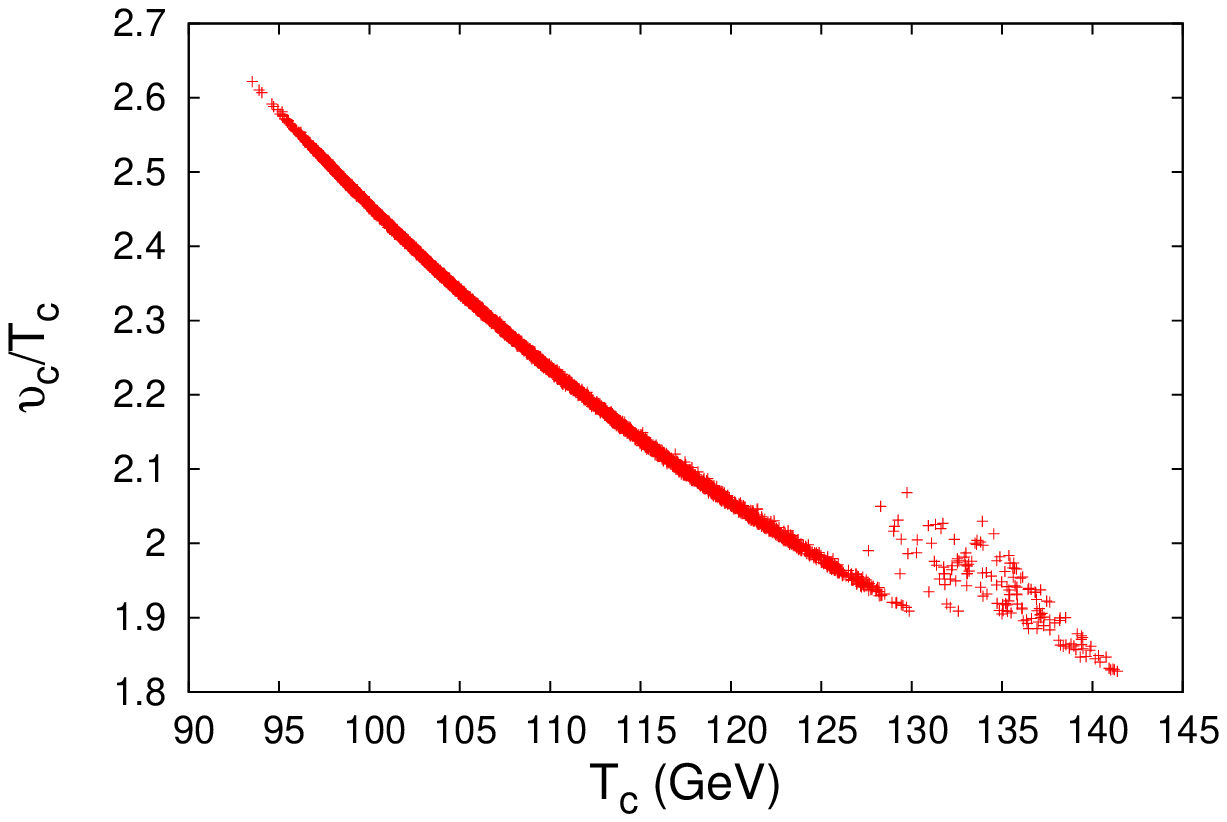}
\includegraphics[width=8cm,height=6cm]{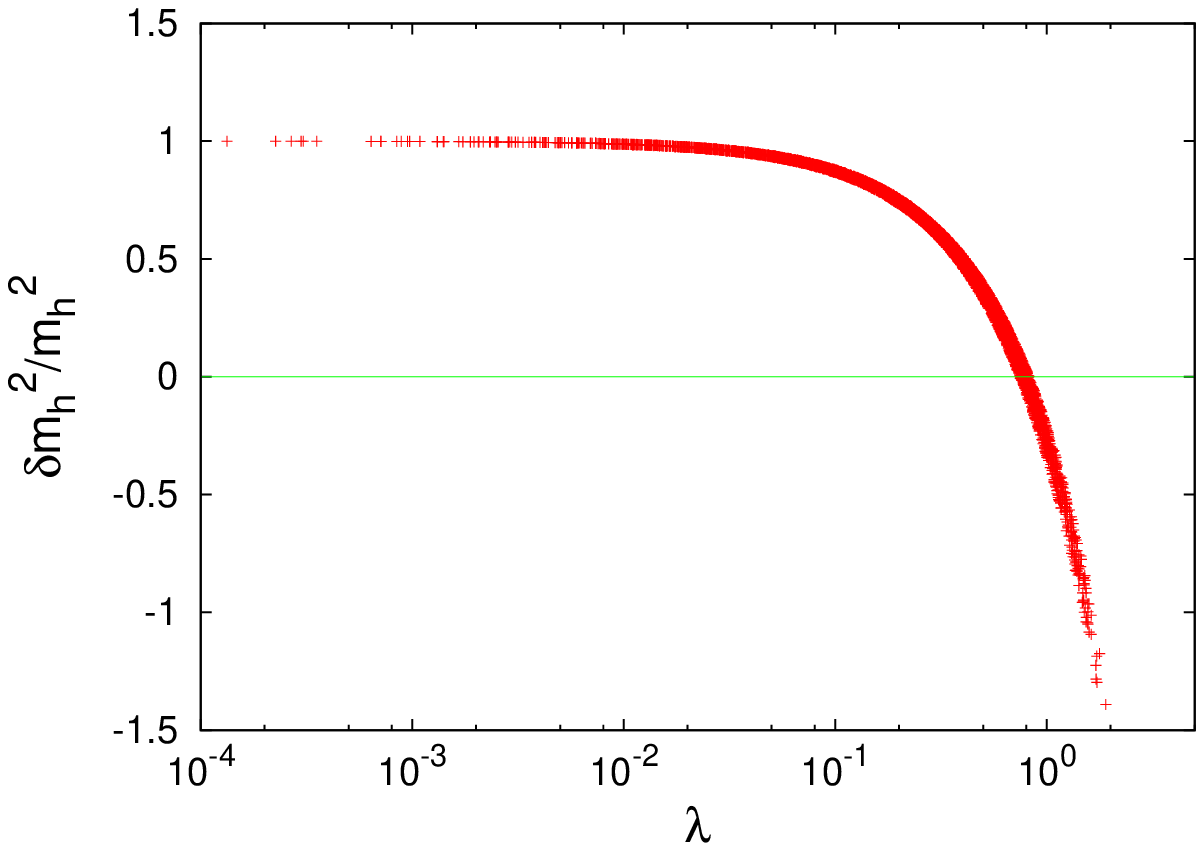}\caption{\textit{In the left
figure, the critical temperature is presented versus the quantity
$\upsilon_{c}/T_{c}$ in (\ref{v/t}). In the right one, the relative
contribution of the one-loop corrections (including the counter-terms) to the
Higgs mass versus the parameter $\lambda$.}}%
\label{ct}%
\end{figure}

In Fig. \ref{ct}, we show two plots: one for $\upsilon(T_{c})/T_{c}$ versus
the critical temperature, and the second one for the dependence of the one
loop correction to the Higgs mass on its quartic coupling for the same sets of
parameters used in Fig. \ref{Hr}\ in the previous section. It is worth noting
that the parameters $\eta_{1}$, $\eta_{2}$ and $\eta_{12}$ in (\ref{nu-mass})
do not play a significant role in the dynamics of the EWPT, and therefore we
fixed them in such a way to avoid the existence of electric charge breaking minima.

From the left panel in Fig. \ref{ct}, we can see that one can have a strongly
first order EWPT while the critical temperature lies around 100 $\mathrm{GeV}%
$. The right panel shows that the one-loop contribution to the Higgs mass can
be large compared to its tree-level value for small values of the self
coupling $\lambda$. For larger values of $\lambda$, this contribution can be
negative in order to bring the large tree-level Higgs mass down to its
experimental value. Therefore, the EWPT can easily be strongly first order
without being in conflict with the measured value of the Higgs mass.

\begin{figure}[t]
\begin{center}
\includegraphics[width=8cm,height=6cm]{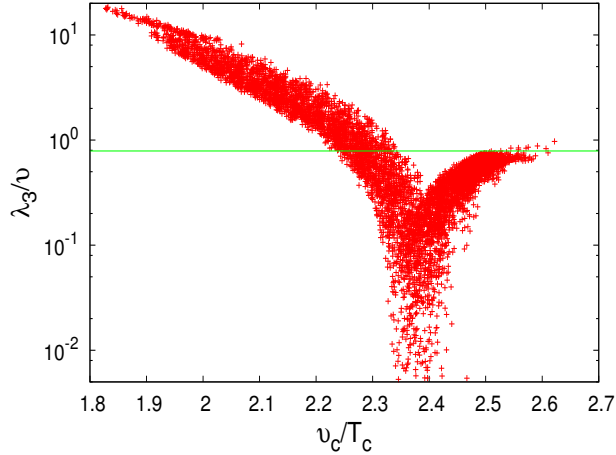}
\end{center}
\caption{\textit{The triple Higgs coupling $\lambda_{3}$ in absolute value
estimated at one loop in units of the Higgs vev; is shown versus the quantity
$\upsilon_{c}/T_{c}$ in (\ref{v/t}). The green line represents the tree-level
value, and the corresponding benchmarks are the cases where different one-loop
corrections cancel each other.}}%
\label{hhh}%
\end{figure}

Another issue in the investigation of the EWPT that could have impacts on
collider signatures is the possible connection between the EWPT strength and
the value of the mass-dimension triple Higgs coupling $\lambda_{3}$ as first
discussed in \cite{KOS}. In order to show this correlation between the EWPT
strength and the enhancement on the triple Higgs coupling due to the
non-decoupling loop effect of the additional charged scalars, we use the same
values of the parameters of Fig. \ref{ct}; and plot the triple Higgs coupling
scaled the Higgs vev versus the EWPT strength, i.e., the ratio (\ref{v/t}) as
shown in Fig. \ref{hhh}.

It is clear that the triple Higgs coupling one-loop corrections could be very
large with respect to the tree-level value for $\upsilon_{c}/T_{c}\lesssim2.2$.

According to the ILC physics subgroup, the triple Higgs coupling can be
measured with about 20\% accuracy or better at $\sqrt{s}=500$ \textrm{GeV}
with integrated luminosity $\mathcal{L}=500$ $fb^{-1}$ \cite{ILCnew}. This
implies that for large parameter space, the model can be potentially testable
at future linear colliders.

\section{Collider Phenomenology}

Since the RH neutrinos couple to the charged leptons, one excepts them to be
produced at $e^{-}e^{+}$ colliders, such as the ILC and CLIC with a collision
energy $\sqrt{s}$ of few hundreds \textrm{GeV} up to TeV. If the produced
pairs are of the form $N_{2,3}N_{2,3}$ or $N_{1}N_{2,3}$, then $N_{2,3}$ will
decay into charged lepton and $S_{2}^{\pm}$, and subsequently $S_{2}^{\pm}$
will decay into $N_{1}$ and a charged lepton. If such decays occur inside the
detector, then the signal will be
\[
\left\{
\begin{array}
[c]{ccc}%
\not E +2\ell_{R}, & \text{for} & e^{+}e^{-}\rightarrow N_{1}N_{2,3}\\
\not E +4\ell_{R}, & \text{for} & e^{+}e^{-}\rightarrow N_{2,3}N_{2,3}.
\end{array}
\right.
\]
However, for $m_{N_{i}}\geq100~$\textrm{GeV}, it is very possible that the
decay $N_{2,3}\rightarrow N_{1}+2\ell_{R}$ occurs outside the detector, and
thus escapes the detection. In this section, we assume that this is the case.
Therefore, we analyze the production of all possible pairs of RH neutrinos,
tagged with a photon from an initial state radiation, that is $e^{-}%
e^{+}\rightarrow N_{i}N_{k}\gamma$ (with $i,k=1,2,3$), where one searches for
a high $p_{T}$ gamma balancing the invisible RH neutrinos.

If the emitted photon is soft or collinear, then one can use the
soft/collinear factorization form \cite{Birkedal}
\begin{equation}
\frac{d\sigma\left(  e^{+}e^{-}\rightarrow N_{i}N_{k}\gamma\right)  }%
{dxd\cos\theta}\simeq\mathcal{F}(x,\cos\theta)\hat{\sigma}\left(  e^{+}%
e^{-}\rightarrow N_{i}N_{k}\right)  , \label{fsr}%
\end{equation}
with $x=2E_{\gamma}/\sqrt{s}$, here $\theta$\ is the angle between the photon
and electron and $\hat{\sigma}$ is the cross section (\ref{cross}) evaluated
at the reduced center of mass energy $\hat{s}=(1-x)s$. The function
$\mathcal{F}$\ has a universal form
\begin{equation}
\mathcal{F}(x,\cos\theta)=\frac{\alpha_{em}}{\pi}\frac{1+(1-x)^{2}}{x}\frac
{1}{\sin^{2}\theta}.
\end{equation}
Collinear photon with the incident electron or positron could be a good
positive signal, especially if the enhancement in (\ref{fsr}) is more
significant than the SM background.

There are two leading SM background processes: a) the neutrino counting
process $e^{-}e^{+}\rightarrow\nu{\bar{\nu}}\gamma$ from the t-channel $W$
exchange and the s-channel $Z$ exchange, and b) the Bhabha scattering with an
extra photon $e^{-}e^{+}\rightarrow e^{-}e^{+}\gamma$, which can mimic the
$N_{i}N_{i}$ signature when the accompanying electrons or photons leave the
detector through the beam pipe \cite{BOUJ}. In addition to putting the cut on
the energy of the emitted photon, one can reduce further the mono-photon
neutrino background, by polarizing the incident electron and positron beams
such that
\begin{equation}
\frac{N_{e_{R}^{-}}-N_{e_{L}^{-}}}{N_{e_{R}^{-}}+N_{e_{L}^{-}}}%
>>50\%;~~~~~~~\frac{N_{e_{R}^{+}}-N_{e_{L}^{+}}}{N_{e_{R}^{+}}+N_{e_{L}^{+}}%
}<<50\%,
\end{equation}
where $N_{e_{L,R}^{-}}$ and $N_{e_{L,R}^{+}}$ are the number densities of the
left (right)-handed electrons and positrons per unit time in the beam. At
$\sqrt{s}>>100$ \textrm{GeV} the process $e^{-}e^{+}\rightarrow\nu{\bar{\nu}%
}\gamma$ is dominated by the $W$-exchange, and hence one expect that having
the electron (positron) beam composed mostly of polarized right handed (left
handed) electrons (positron) reduces this background substantially, whereas
the signal increases since $N_{i}$ couples to the right handed electrons.

Now, let us estimate the total cross section $\sigma\left(  e^{+}%
e^{-}\rightarrow N_{i}N_{k}\right)  $, which is basically the reverse of one
of the processes which determines the effective dark matter density for
coannihilation, at a collision energy of $\sqrt{s}$. The differential cross
section of $e^{+}e^{-}\rightarrow N_{i}N_{k}$ for the energy $\sqrt{s}$ is
given by \cite{Keung}
\begin{equation}
\tfrac{d\sigma\left(  e^{+}e^{-}\rightarrow N_{i}N_{k}\right)  }{d\cos\theta
}=\kappa_{ik}\tfrac{\left\vert g_{ie}g_{ke}^{\ast}\right\vert ^{2}}{128\pi
}\tfrac{\beta_{ik}}{s}\left(  \tfrac{\left(  \tilde{t}-x_{i}\right)  \left(
\tilde{t}-x_{k}\right)  }{\left(  \tilde{t}-x_{s}\right)  ^{2}}+\tfrac{\left(
\tilde{u}-x_{i}\right)  \left(  \tilde{u}-x_{k}\right)  }{\left(  \tilde
{u}-x_{s}\right)  ^{2}}-\tfrac{2\sqrt{x_{i}x_{k}}}{\left(  \tilde{t}%
-x_{s}\right)  \left(  \tilde{u}-x_{s}\right)  }\right)  ,
\end{equation}
with $\kappa_{ik}=1/2$ if the two RH neutrinos are identical and equal to one
if they are different, $\theta$\ is the angle between the incoming electron
and the outgoing $N_{i}$, and
\begin{align}
x_{j}  &  =m_{Nj}^{2}/s,~x_{s}=m_{S_{2}}^{2}/s,~\beta_{ik}=\sqrt{\left(
1-x_{i}-x_{k}\right)  ^{2}-4x_{i}x_{k}}\nonumber\\
\tilde{t}  &  =\frac{t}{s}=\tfrac{x_{i}+x_{k}}{2}-\tfrac{1}{2}\left(
1-\beta_{ik}\cos\theta\right)  ,~\tilde{u}=\frac{u}{s}=\tfrac{x_{i}+x_{k}}%
{2}-\tfrac{1}{2}\left(  1+\beta_{ik}\cos\theta\right)  ,
\end{align}
By integrating over $\cos{\theta}$, the total cross section reads
\begin{equation}
\sigma\left(  e^{+}e^{-}\rightarrow N_{i}N_{k}\right)  =\kappa_{ik}
\tfrac{\left\vert g_{ie}g_{ke}^{\ast}\right\vert ^{2}}{32\pi}\tfrac{\beta
_{ik}}{s}\left\{  1+\tfrac{4\left[  x_{s}^{2}-x_{i}x_{s}-x_{k}x_{s}+x_{i}%
x_{k}\right]  }{w^{2}-\beta_{ik}^{2}}+\tfrac{w^{2}+w+2\sqrt{x_{i}x_{k}}}%
{\beta_{ik}w}\ln\left(  \tfrac{w-\beta_{ik}}{w+\beta_{ik}}\right)  \right\}  ,
\label{cross}%
\end{equation}
with $w=-1+x_{i}+x_{k}-2x_{s}$. In order to estimate the differential cross
section of the process $e^{+}e^{-}\rightarrow N_{i}N_{k}\gamma$ we integrate
(\ref{fsr}) over $\theta$ taking into account the minimum value of
electromagnetic calorimeter acceptance in the ILC to be $\sin\theta
>0.1$\ \cite{ILC}.

\begin{figure}[t]
\begin{centering}
\includegraphics[width=8cm,height=6.5cm]{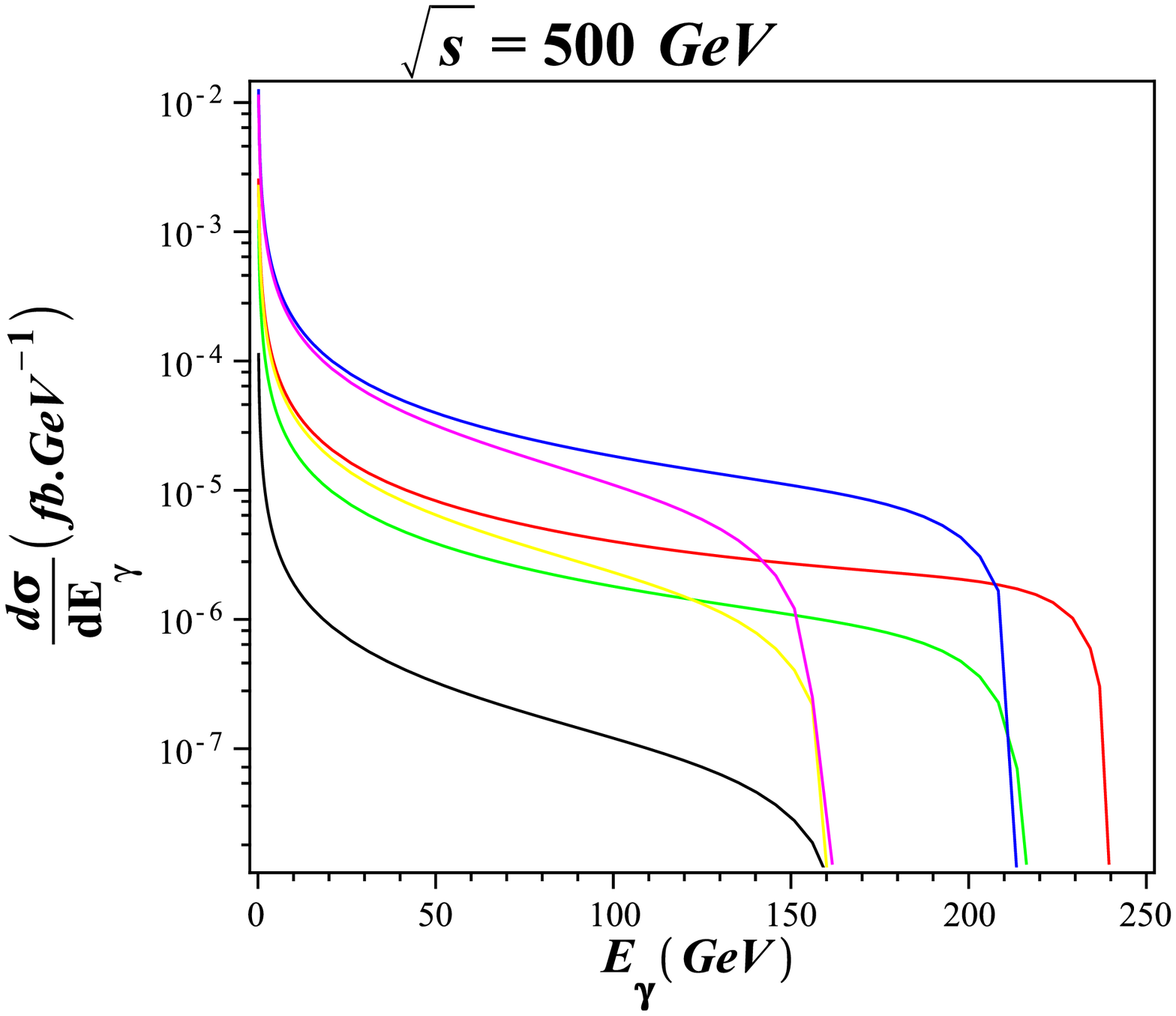}\includegraphics[width=8cm,height=6.5cm]{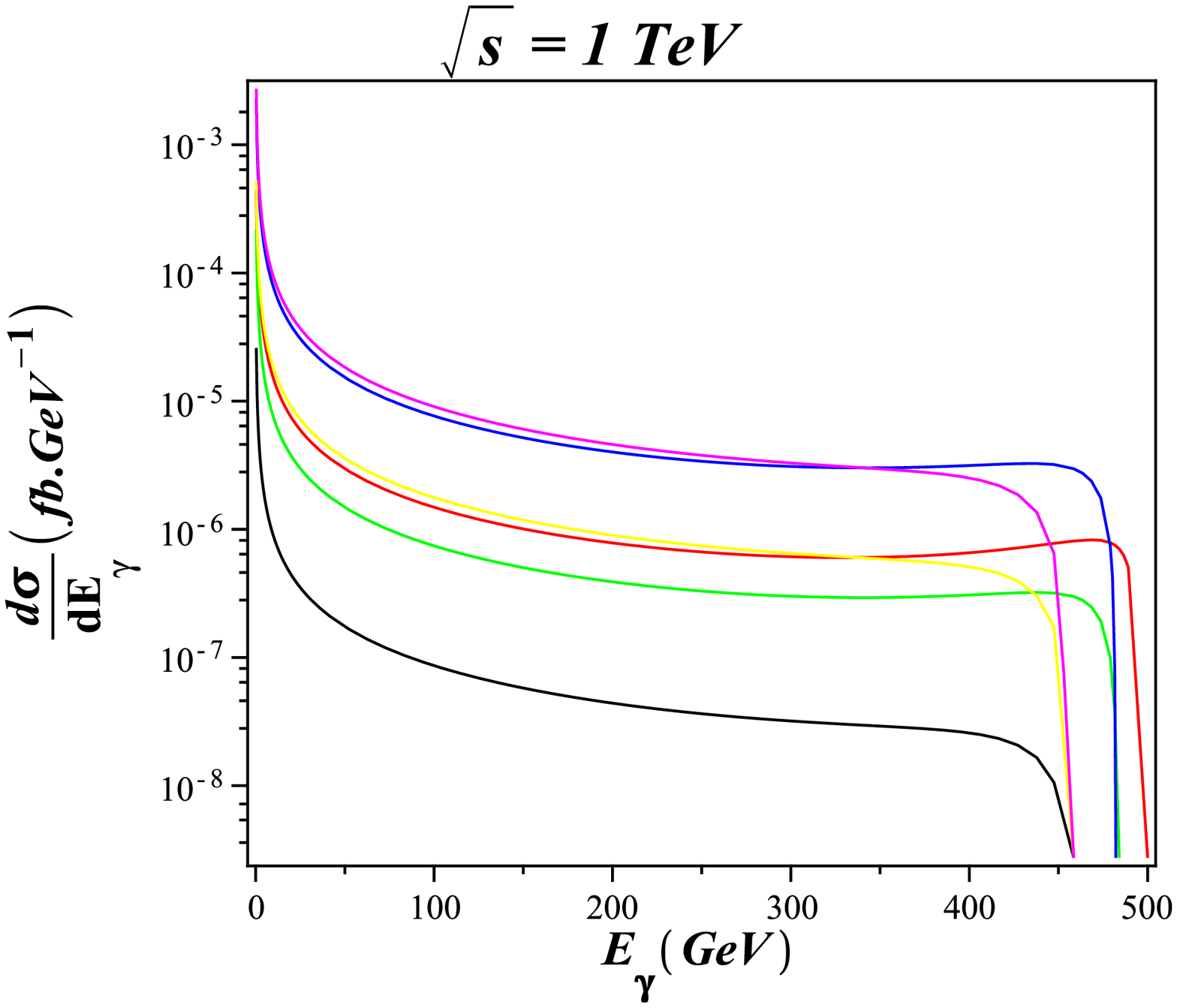}
\par\end{centering}
\caption{\textit{The photon spectra from the processes $e^{+}e^{-}\rightarrow
N_{i}N_{k}\gamma$ where the curves: {red, green, black, blue, yellow, magenta}
correspond to (i,k)={(1,1), (1,2), (2,2), (1,3), (2,3), (3,3)} respectively.
Here, we considered the following favored mass values: $m_{N_{1}}=52.53$ GeV,
$m_{N_{2}}=121.80$ GeV, $m_{N_{3}}=126.19$ GeV, $m_{S_{2}}=144.28$ GeV, and
the coupling values: $g_{1e}=-4.19\times10^{-2}$, $g_{2e}=2.10\times10^{-2}$
and $g_{3e}=-6.75\times10^{-2}$.}}%
\label{Ga}%
\end{figure}In Fig. \ref{Ga}, we show the photon spectrum for two values of
collision energies $\sqrt{s}=500$\ \textrm{GeV}\ and 1 \textrm{TeV}. These
plots are estimated using the factorization formula (\ref{fsr}), however, we
obtain similar results using CalcHEP \cite{CalcHEP}. We see that for the
benchmark shown in Fig. \ref{Ga}, the heaviest RH neutrino is largely produced
due to its large couplings to the electron/positron. Thus, for this particular
benchmark the missing energy is dominated not by the DM, but rather by the
other RH neutrinos.

Another interesting process that might be possible to search for at both
lepton and hadron colliders is the production of $S_{1,2}^{\pm}$. For
instance, at the LHC they can be pair produced in an equal number via the
Drell-Yan process, with the partonic cross section at the leading order given
by
\begin{align}
{\hat\sigma} = \frac{\pi\alpha^{2}Q^{2}_{q}}{3 {\hat s}}%
\end{align}
where ${\hat s}$ is the energy squared in the center of mass frame of the
quarks, and $Q_{q}$ stands for the parton's electric charge. Thus, from the
dependence on the energy of the partons, we see that the production rate of
$S_{1,2}^{\pm}$ is suppressed at very high energies, and so we expect that
most of the produced $S_{1,2}^{\pm}$ will have energies not too far from their
masses. Now, Each pair of charged scalars decays into charged leptons and
missing energy, such as $e^{+}e^{-}$, $\mu^{+}\mu^{-}$, $\mu^{+}e^{-}$. The
observation of an electron (positron) and anti-muon (muon), will be a strong
signal for the production of the charged scalars of this model. The energy
carried out by the charged leptons, $\ell_{\alpha}^{+}\ell_{\beta}^{-}$,
produced in the decay of $S_{1,2}^{\pm}$ will be limited by the phase space
available to $N_{1}$ and $\ell_{\alpha,\beta}$ since $m_{S_{2}} - m_{N_{1}} <<
m_{S_{2}}$. On the other hand, the leptons originating from the decay of
$S_{1}^{\pm}$ will be produced in association with a SM neutrino, and hence
can have energy as large as $\sim m_{S_{1}}$. Thus, by putting the appropriate
energy cuts on the energy of final states $e^{\pm} \mu^{\mp}$ and
discriminating the SM background (from the decay of $p p \rightarrow
W^{+}W^{-} + X \rightarrow\ell_{\alpha}\ell_{\beta} + \nu^{\prime}s $), one
can, in principle, identify the signal for the charged scalars. This requires
a detailed a analysis which we plan to carry out in a future publication
\cite{ANS}.

\section{Conclusion}

In this paper, we analyzed a radiative model for neutrino masses, generated at
three loop level. Beside it can accommodate the neutrino oscillation data and
be consistent with the LFV processes, it provides a DM candidate with a mass
lying between few \textrm{GeV} up to 225 \textrm{GeV}; and a relatively light
charged scalar, $S_{2}^{\pm}$, with a mass below 245 \textrm{GeV}.
Furthermore, we showed that the charged scalar singlets can give an
enhancement for $B\left(  h\rightarrow\gamma\gamma\right)  $, whereas the
decay $B\left(  h\rightarrow\gamma Z\right)  $ get a small suppression,
compared to the SM. We also found that charged scalars with masses close the
electroweak scale make the electroweak phase transition strongly first order.
Since $N_{1}$ couples only to leptons, it can not be observed in experiments
for direct dark matter searches. However it might be possible to search for
such particle in indirect detection experiments, such as Fermi-LAT, and at
future linear colliders, such as the international linear collider (ILC).
Thus, for this particular benchmark the missing energy is dominated not by the
DM, but rather by the other RH neutrinos.

\begin{acknowledgments}
We would like to thank S. Kanemura for his careful reading and
useful comments on the manuscript. We also thank R. Soualah for the
useful discussion on the search for DM at collider. This work is
supported by the Algerian Ministry of Higher Education and
Scientific Research under the PNR '\textit{Particle
Physics/Cosmology: the interface}', and the CNEPRU Projects No.
\textit{D01720130042} and \textit{D01720090023}.
\end{acknowledgments}

\appendix

\section{Exact Neutrino Mass}

According to Fig. \ref{nu-mass}, the neutrino mass matrix element
[$\alpha,\beta$] is given by
\begin{align}
(M_{\nu})_{\alpha\beta}  &  =2\int\tfrac{d^{4}Q_{1}}{\left(  2\pi\right)
^{4}}\int\tfrac{d^{4}Q_{2}}{\left(  2\pi\right)  ^{4}}\tfrac{i}{Q_{1}%
^{2}-m_{S_{1}}^{2}}\left(  -2if_{\alpha i}\right)  P_{L}\tfrac{i}{\not Q
_{1}-m_{\ell_{i}}}\times\nonumber\\
&  i\Gamma_{ik}\left(  Q_{1},Q_{2},-Q_{1},-Q_{2}\right)  \tfrac{i}{\not Q
_{2}-m_{\ell_{k}}}P_{L}\left(  -2if_{\beta k}\right)  \tfrac{i}{Q_{2}%
^{2}-m_{S_{1}}^{2}},
\end{align}
with $P_{L,R}=(1\mp\gamma_{5})/2$, and the effective vertex $\Gamma_{ik}%
(p_{1},p_{2},k_{1},k_{2})$\ is a function of the momenta $p_{1,2}$ and
$k_{1,2}$ for charged leptons and scalars, respectively, which is given by
\begin{equation}
i\Gamma_{ik}\left(  p_{1},p_{2},k_{1},k_{2}\right)  =\int\tfrac{d^{4}%
k}{\left(  2\pi\right)  ^{4}}\left(  -ig_{ij}\right)  P_{L}\tfrac{i\left(
\not k  +m_{N_{j}}\right)  }{k^{2}-m_{N_{j}}^{2}}P_{L}\left(  -ig_{kj}\right)
\tfrac{i}{\left(  k+p_{1}\right)  ^{2}-m_{S_{2}}^{2}}\left(  -i\lambda
_{s}\right)  \tfrac{i}{\left(  k-p_{2}\right)  ^{2}-m_{S_{2}}^{2}}.
\end{equation}
Then the neutrino mass matrix element [$\alpha,\beta$] is
\begin{align}
(M_{\nu})_{\alpha\beta}  &  =8f_{\alpha i}f_{\beta k}m_{\ell_{i}}m_{\ell_{k}%
}\lambda_{s}g_{ij}g_{kj}m_{N_{j}}P_{L}\int\tfrac{d^{4}k}{\left(  2\pi\right)
^{4}}\tfrac{1}{\left(  k^{2}-m_{N_{j}}^{2}\right)  }\times\nonumber\\
&  \int\tfrac{d^{4}Q_{1}}{\left(  2\pi\right)  ^{4}}\tfrac{1}{\left(
Q_{1}^{2}-m_{\ell_{i}}^{2}\right)  \left(  Q_{1}^{2}-m_{S_{1}}^{2}\right)
\left(  \left(  k+Q_{1}\right)  ^{2}-m_{S_{2}}^{2}\right)  }\int\tfrac
{d^{4}Q_{2}}{\left(  2\pi\right)  ^{4}}\tfrac{1}{\left(  Q_{2}^{2}-m_{\ell
_{k}}^{2}\right)  \left(  Q_{2}^{2}-m_{S_{1}}^{2}\right)  \left(  \left(
k+Q_{2}\right)  ^{2}-m_{S_{2}}^{2}\right)  },
\end{align}
and since we have
\begin{equation}
\tfrac{1}{\left(  Q^{2}-m_{0}^{2}\right)  \left(  Q^{2}-m_{1}^{2}\right)
}=\tfrac{1}{m_{1}^{2}-m_{0}^{2}}\left(  \tfrac{1}{Q^{2}-m_{1}^{2}}-\tfrac
{1}{Q^{2}-m_{0}^{2}}\right)  ,
\end{equation}
thus
\begin{equation}
\int\tfrac{d^{4}Q}{\left(  2\pi\right)  ^{4}}\tfrac{1}{\left(  Q^{2}-m_{0}%
^{2}\right)  \left(  Q^{2}-m_{1}^{2}\right)  \left(  \left(  k+Q\right)
^{2}-m_{2}^{2}\right)  }=\frac{i\left(  B_{0}\left(  k^{2},m_{1}^{2},m_{2}%
^{2}\right)  -B_{0}\left(  k^{2},m_{0}^{2},m_{2}^{2}\right)  \right)  }%
{16\pi^{2}m_{1}^{2}},
\end{equation}
where the $B_{0}$ Passarino-Veltman function is \cite{PV}
\begin{equation}
B_{0}\left(  k^{2},m_{1}^{2},m_{2}^{2}\right)  =\frac{1}{\epsilon}-\int
_{0}^{1}dx\ln\tfrac{-x\left(  1-x\right)  k^{2}+\left(  1-x\right)  m_{1}%
^{2}+xm_{2}^{2}}{\mu^{2}}.
\end{equation}
Then, by neglecting the charged lepton masses, and making Wick rotation, we
get%
\begin{equation}
(M_{\nu})_{\alpha\beta}=i\tfrac{\lambda_{s}m_{\ell_{i}}m_{\ell_{k}}}{\left(
4\pi^{2}\right)  ^{3}m_{S_{2}}}f_{\alpha i}f_{\beta k}g_{ij}g_{kj}F\left(
m_{N_{j}}^{2}/m_{S_{2}}^{2},m_{S_{1}}^{2}/m_{S_{2}}^{2}\right)  ,
\end{equation}
with%
\begin{equation}
F\left(  \alpha,\beta\right)  =\frac{\sqrt{\alpha}}{8\beta^{2}}\int
_{0}^{\infty}dr\tfrac{r}{r+\alpha}\left(  \int_{0}^{1}dx\ln\tfrac{x\left(
1-x\right)  r+\left(  1-x\right)  \beta+x}{x\left(  1-x\right)  r+x}\right)
^{2}. \label{FF}%
\end{equation}

\section{Thermal Masses}

The thermal masses are given as $\tilde{m}_{i}^{2}(h)=m_{i}^{2}(h)+\Pi
_{i}\left(  T\right)  $, where $m_{i}^{2}(h)$ and $\Pi_{i}\left(  T\right)  $
are the field-dependant masses and the thermal self-energies respectively. In
this model, the field-dependant masses are given by%
\begin{align}
m_{W}^{2}  &  =g_{2}^{2}\tfrac{h^{2}}{4},~m_{t}^{2}=y_{t}^{2}\tfrac{h^{2}}%
{2},~m_{\chi}^{2}=-\mu^{2}+\lambda\tfrac{h^{2}}{6},~m_{h}^{2}=-\mu^{2}%
+\lambda\tfrac{h^{2}}{2},\nonumber\\
~m_{S_{i}}^{2}  &  =m_{i}^{2}+\lambda_{i}\tfrac{h^{2}}{2},~m_{W^{3}}%
^{2}(h)=g_{2}^{2}\tfrac{h^{2}}{4},~m_{W^{3}-B}^{2}(h)=g_{2}g_{1}\tfrac{h^{2}%
}{4},~m_{B}^{2}(h)=g_{1}^{2}\tfrac{h^{2}}{4}, \label{fm1}%
\end{align}
where the diagonalization of the $\{W^{3}-B\}$ matrix gives the massless
photon and $Z$ mass $m_{Z}^{2}=\left(  g_{2}^{2}+g_{1}^{2}\right)  h^{2}/4$.
The thermal self-energies, that are generally estimated in the high
temperature approximated as $\Pi_{i}\sim T^{2}$. In this model, these thermal
self-energies are given by%
\begin{align}
\Pi_{h}  &  =\Pi_{\chi}=T^{2}\left\{  \frac{\lambda}{12}+\frac{3g_{2}%
^{2}+g_{1}^{2}}{16}+\frac{y_{t}^{2}}{4}+\frac{\lambda_{1}}{6}+\frac
{\lambda_{2}}{6}\right\}  ,\nonumber\\
\Pi_{W^{\pm}}  &  =\Pi_{W^{3}}=\frac{11}{6}g_{2}^{2}T^{2},~\Pi_{W^{3}%
-B}=0,\nonumber\\
\Pi_{B}  &  =\frac{11}{6}g_{1}^{2}T^{2}+\frac{1}{6}g_{1}^{2}T^{2}+\frac{1}%
{6}g_{1}^{2}T^{2}.\nonumber\\
\Pi_{S_{1}}  &  =T^{2}\left\{  \frac{\lambda_{1}}{3}+\frac{\eta_{1}}{6}%
+\frac{\eta_{12}}{6}\right\}  ,\\
\Pi_{S_{2}}  &  =T^{2}\left\{  \frac{\lambda_{2}}{3}+\frac{\eta_{12}}{6}%
+\frac{\eta_{2}}{6}\right\}  . \nonumber\label{ThM}%
\end{align}
The last two terms in $\Pi_{h}$, $\Pi_{\chi}$, $\Pi_{B}$, $\Pi_{S_{1}}$ and
$\Pi_{S_{2}}$ represent the thermal loop contributions of $S_{1}^{\pm}$ and
$S_{2}^{\pm}$ respectively.

\end{document}